\documentclass{emulateapj}
\usepackage{psfig}
\usepackage{graphicx}

\submitted{Submitted to The Astrophysical Journal (April 2,
2004)}

\advance \voffset by 0.00cm\relax
\def\msun{{\rm ~M}_{\odot}}

\shorttitle{Young Black Hole Populations}
\shortauthors{Belczynski, Sadowski \& Rasio}

\begin{document}

\title{A Comprehensive Study of Young Black Hole Populations}

\author{Krzysztof Belczynski\altaffilmark{1,2}, Aleksander 
        Sadowski\altaffilmark{3}, Frederic A.\ Rasio\altaffilmark{1}}

\affil{
     $^{1}$ Northwestern University, Dept of Physics and Astronomy,
       2145 Sheridan Rd, Evanston, IL 60208\\
     $^{2}$ Lindheimer Postdoctoral Fellow\\  
     $^{3}$ Warsaw University, Dept. of Physics, 
            Hoza 69, 00-681, Warsaw, Poland\\
     belczynski, rasio@northwestern.edu; oleks@camk.edu.pl}

\begin{abstract} 
We present theoretical models of black hole (BH) populations 
in young stellar environments, such as starbursts and young star clusters. 
Using a population synthesis approach we compute the formation rates 
and characteristic properties of single and binary BHs for various
representative ages and choices of parameters.
We find that most of the BHs (typically $80\%$ for an initial 50\% 
binary fraction) are single, but with many originating 
from primordial binaries (which either merged into a single massive star 
or were disrupted following a supernova explosion).
A smaller but significant fraction (typically $20\%$) of the BHs 
remain in binary systems.
Main-sequence stars are the most frequent BH companions, but massive 
BH--BH binaries are the next most numerous group.
The most massive BHs found in our simulations reach $\sim 80\,M_\odot$, 
and are formed through mergers of massive binary components. 
If formed in a dense star cluster such a massive stellar BH may become 
the seed for growth to a more massive (``intermediate-mass'') BH.
Although we do not include dynamical interactions, our results  
provide realistic initial conditions for $N$-body
simulations of dense star clusters (e.g., globular clusters) including
primordial BHs.
\end{abstract}


\section{INTRODUCTION}

Until recently it was believed that BHs existed in two 
separate mass ranges: stellar-mass BHs, with masses $\sim 10\,M_\odot$, 
and supermassive BHs, with masses $\sim 10^6 - 10^9\,M_\odot$ 
(see, e.g., Fryer \& Kalogera 2001; Peterson 2003; Tegmark 2002). 
However, over 
the last few years, evidence has been mounting for the existence of
{\em intermediate-mass\/} BHs (Miller \& Colbert 2004). 
Although the observations remain controversial 
a variety of plausible formation scenarios for intermediate-mass BHs (IMBHs)
have been proposed (e.g., van der Marel 2003).    
Primordial formation of massive BHs in the early universe has been discussed
for many years (e.g., Carr 1993; Khlopov, Rubin \& Sakharov 2002). 
Later on, when the first very massive metal-free (Pop~III) stars form, 
one naturally expects that, given the lack of significant 
wind mass loss, these stars may collapse to form IMBH remnants with
masses up to $\sim 10^2 - 10^3\, M_\odot$ 
(e.g., Heger et al.\ 2003). Some of these IMBHs may be in binaries
either formed through captures in dense environments (Wyithe \& Loeb 2003) 
or remaining from primordial Pop~III binaries (Belczynski, Bulik \& Rudak 2004a).    
Dynamical formation processes for IMBHs in Pop~II and Pop~I star clusters
have also been the subject of many recent studies. The two most promising
scenarios involve
runaway collisions and mergers of massive main-sequence stars 
(Portegies Zwart \& McMillan 2002; G{\"u}rkan, Freitag, \& Rasio 2004)
and successive mergers of stellar-mass BH (Miller \& Hamilton 2002). 
In this paper, we focus on the populations of stellar BHs forming  
out of the most massive stars in Pop~II or Pop~I young stellar environments. 

Stellar BHs form through core collapse of massive stars with masses 
$\ga 20\,M_\odot$. If $N$ stars form with a standard Kroupa (Kroupa \&
Weidner 2003) initial mass function (IMF) between $M_{\rm min} = 0.08\, 
M_\odot$ and $M_{\rm max} = 150\, M_\odot$, we expect $N_{\rm BH} \simeq 
5 \times 10^{-4}\, N$ black holes to have formed after $\sim 10^7\,$yr, 
when all stars massive enough to produce a BH have evolved. 
The goal of this paper is to characterize this initial population of BHs. 
For a population of {\em single stars\/}, this is merely a question of 
characterizing the metallicity-dependent relation between progenitor 
mass and final BH mass (Fryer \& Kalogera 2001; Heger et al.\ 2003). 
However, most stars, and, especially, most massive stars are expected 
to form in binary systems (Duquennoy \& Mayor 1991). The BH formation
processes can be affected significantly by the evolution of the progenitor
star in a binary, especially if the initial orbital separation is $\la
500\,R_\odot$ and the star can overflow its Roche lobe as it evolves. In
addition, many BHs will retain binary companions, which can drastically
affect their later evolution, their detectability as X-ray or 
gravitational-wave sources, and the way they interact with their environment.

The study we present in this paper is based on a population synthesis
approach, in which a large number of single and binary stars are evolved
according to parametrized evolutionary prescriptions. 
Stellar evolution is followed starting from a population of zero-age 
main-sequence (ZAMS) stars with specified metallicity. Given enough time the 
stars evolve to form remnants, either white dwarfs (not relevant for the massive
stars considered here), neutron stars, or BHs.
Both observations and more detailed theoretical calculations are
used to constrain and parametrize the uncertain stages of evolution 
(e.g., supernova explosions and common envelope phases). 
A number of population synthesis codes of varying levels of
sophistication are currently being used to study many astrophysical problems
(e.g., Fryer, Woosley, \& Hartmann 1999; Nelemans, Yungelson \& 
Portegies-Zwart 2001; Hurley, Tout \& Pols 2002; Pfahl, Rappaport, \& 
Podsiadlowski 2003; Belczynski, Kalogera \& Bulik 2002, hereinafter BKB02).
Here we use the most recent version of the {\tt StarTrack} population synthesis 
code (BKB02; Belczynski et al. 2004b) to study BH formation 
in young metal-rich (Pop~I) and metal-poor (Pop~II) populations. This work
is complementary to the previous study of the oldest binary BHs, descendants of 
metal-free Pop~III stars, by Belczynski et al.\ (2004a).

Our results can be applied directly to the modeling of young starburst 
populations, with ages $\sim 10^7-10^8\,$yr. For those systems, we have derived 
complete synthetic sample catalogues of all BHs, single and in binaries, that have 
evolved from the initial burst of star formation. Our results can also be 
applied to the modeling of young star clusters (such as the super star 
clusters so prevalent in starbursts; see, e.g., de Grijs et al. 2003), 
although we do not take into account the effects of dynamical interactions 
in dense cluster cores (cf.\ G\"urkan, Freitag \& Rasio 2003; Ivanova et 
al.\ 2004).

Characterizing the properties of a primordial BH population is also very
important for many theoretical studies of old star clusters. Indeed many of
these studies attempt to bypass the first $\sim 10^7$yr of massive star
evolution and use initial conditions that already contain BHs. For example,
$N$-body simulations of globular clusters containing primordial BHs have 
been performed starting with a small number of identical single BHs of mass
$10\,M_\odot$ (e.g., Fregeau et al.\ 2002; Portegies Zwart \& McMillan
2000). One goal of our work is to provide more realistic initial conditions
for those studies. The further dynamical evolution of BHs in dense star
clusters could play a key role in many problems of great current 
interest, such as understanding the
formation of IMBHs and ultra-luminous X-ray sources (Miller \& Hamilton 2002) 
and predicting the
merger rate of BH binaries detectable by gravitational-wave detectors
(Portegies Zwart \& McMillan 2000).

Our paper is organized as follows. In \S\,2 we describe the evolutionary
model used in the population synthesis calculations, including  
details about BH formation. In \S\,3 we present the results of our
calculations. First we present the 
results for our reference model, then we follow with the description of a 
number of alternative models. 
Finally, in \S\,4 we summarize our main results.

\section{MODEL DESCRIPTION AND ASSUMPTIONS}

\subsection{Population Synthesis Code}

Our calculations are based on a population synthesis method. 
The {\tt StarTrack} code (BKB02)
has recently undergone major revisions and updates (Belczynski et al.\ 2004b).
These include: a detailed treatment of tidal dissipation effects;
individual treatments of various Roche lobe overflow (RLOF) phases; full 
numerical orbit evolution with angular momentum losses due to magnetic 
braking, gravitational radiation (GR), mass transfer/loss, and tidal
interactions,
and incorporating the stabilizing influence of optically thick winds.

All stars are evolved based on the metallicity-dependent models of 
Hurley, Pols, \& Tout (2000), with the improvements described in BKB02.
Each star, either a single or a binary component, is initiated
on ZAMS, and its evolution is followed from ZAMS through a sequence of 
evolutionary phases: Main Sequence (MS), Hertzsprung Gap (HG),
Red Giant Branch (RG), Core Helium Burning (CHeB), Asymptotic Giant
Branch (AGB), and for stars stripped off their hydrogen-rich layers  
Helium phases (He).
The nuclear evolution of a star ends at the formation of a
stellar remnant: a white dwarf (WD), a neutron star (NS) or a BH.  

We adopt parameters corresponding to the standard model of BKB02 (see 
\S\,2 in BKB02)  incorporating the latest natal kick velocity
distribution of Arzoumanian, Chernoff \& Cordes (2002), and limiting 
the accretion rate onto the NS and the BH at the maximum Eddington limit 
with the rest of the transferred material lost with specific orbital 
angular momentum of the accretor during the dynamically stable RLOF events,
but allowing for hyper-critical accretion during common envelope (CE) phases.
During dynamically stable phases with non-compact accretors, we allow for
non-conservative evolution, with half of the transferred mass lost 
with the specific angular momentum equal to ${2 \pi j A^2 / P}$, where $P$ 
is orbital period, $A$ binary separation, and we choose the scaling
parameter $j=1$ for our calculations (Podsiadlowski, Joss \& Hsu 1992).
In addition to dynamically unstable RLOF events we also allow 
for evolution into the CE phase in cases where the trapping radius 
exceeds the Roche lobe radius of the accretor (e.g., King \& Begelman 1999; 
Ivanova et al. 2003).
Phases preceding CE are driven by angular momentum losses (as described above),  
while the inspiral during CE is treated in the standard manner through a 
prescribed energy efficiency (with $\alpha_{\rm ce} \times \lambda = 1$ in 
our standard model; for details see BKB02).

\subsection{Black Hole Formation}

BHs are formed out of the most massive stars. Using the stellar models 
of Hurley et al.\ (2000) and Woosley (1986) we estimate the time of 
core collapse of the massive star. At that time we know the mass of 
the core (both CO and FeNi core) and the envelope.
For intermediate-mass stars ($\sim 20-30 \msun$ for low metallicity 
models) the FeNi core is collapsed to form a hot proto-NS or a low-mass BH, 
and we use the  work of Fryer (1999) and Fryer \& Kalogera (2001) 
to decide how much fall back is expected in a given case (based 
on the mass of the CO core). Fall-back material is accreted onto the 
central object, increasing its mass, while the rest is assumed to be 
ejected in the supernova (SN) explosion.
For the highest masses ($\gtrsim 30 \msun$ for low metallicity models), 
the entire star goes into collapse, forming 
the BH directly, with no accompanying SN explosion. The regime for 
direct BH formation may be easily seen from Figure~1 (pre-collapse 
mass equal to the remnant mass).

The large observed velocities of radio pulsars imply significant
asymmetries in SN explosions. Although the underlying 
mechanism is not yet understood, it is generally accepted that NS can
receive substantial kicks at birth ($\sim 100-1000\,{\rm km}\,{\rm s}^{-1}$). 
Here we adopt the latest NS kick 
velocity distribution of Arzoumanian et al.\ (2002). Most recent 
observations (e.g., Mirabel \& Rodrigues 2003) suggest that BHs may be 
formed either with an accompanying kick (for smaller mass stellar BHs) 
or without one (for the most massive stellar BHs). 
This folds naturally into our prescription for BH formation.
For low-mass, fall-back BHs, we expect somewhat attenuated SN explosions, 
and we assume that the kick is smaller as well: its magnitude is assumed 
to be inversely proportional to the mass of fall-back material.
For direct BH formation (so called {\em silent\/} collapse) there is 
no explosion, and we therefore assume no asymmetry and no natal kick.

In Figure~1 we show the initial-to-final mass relation for single stars 
for 3 representative metallicities ($Z=$0.02, 0.001, 0.0001). Remnants of various 
types are shown with different symbols. We also plot the instantaneous 
mass of the star just prior to the formation of the compact remnant. 

The top panel of Figure~1 shows the initial-to-final mass relation for solar
metallicity ($Z=0.02$), 
the case described and discussed in detail in BKB02:
BHs are formed from stars more massive than about $20 M_\odot$, with 
a maximum BH mass of $11 M_\odot$.
There is a general rise of BH mass with increasing progenitor mass. 
However, since the wind mass-loss rate increases with the mass of 
the star as well (depleting the mass reservoir for BH formation) the
initial-to-final mass relation flattens out for higher progenitor masses. 
There are two distinctive dips, followed by the subsequent flattening 
of the relation.  
The first (around $25\,M_\odot$), less pronounced dip corresponds to the point 
at which single stars are depleted of their H-rich envelopes, 
entering the Wolf-Rayet stage with enhanced wind mass-loss rates (Hamann \&
Koesterke 1998). Higher 
mass-loss rates reduce the mass of the star and its core, eventually 
leading to the formation of lower-mass remnants.   
The second dip (around $50\,M_\odot$), corresponds to the point where stars 
reach luminosities high enough to initiate the Luminous Blue Variable (LBV) 
phase, characterized by extremely high wind mass-loss rates (Hurley et al.
2000), and a sudden decrease of BH masses.

The middle and bottom panels of Figure~1 show the initial-to-final mass relations 
for lower metallicities ($Z=0.001$ and $Z=0.0001$, respectively).
In each case the relation differs significantly from that obtained for higher
(solar) metallicity. In particular, there is an
increase of the maximum BH mass (to $\sim 27 M_\odot$ for $Z=0.001-0.0001$), 
and the shape of the relation is also altered.
Most of these changes can be attributed to the dependence of the wind 
mass-loss rates on metallicity.
Here we have adopted wind mass-loss prescriptions with square-root dependence on
metallicity ($\propto \sqrt{Z/Z_\odot}$), operating however only during specific 
evolutionary phases (for details see Hurley et al.\ 2000). 
The obvious consequence of smaller mass loss rates for lower
metallicity is the increase of the stellar mass just prior to the
formation of the remnant, leading directly to higher BH masses.
In addition, the smaller mass-loss rates imply that the stars encounter
the LBV dip at a lower mass ($\sim 33 M_\odot$ for $Z=0.001-0.0001$) 
as they evolve to higher luminosities.
The opposite is true for the Wolf-Rayet dip, as now only the stars with much
higher mass lose enough material to become naked helium stars ($\sim 38 
M_\odot$ for $Z=0.001-0.0001$). Note the interesting repositioning of the two 
dips: for solar metallicity the dip at high mass corresponds 
to the LBV phase, while for lower metallicity ($Z=0.001-0.0001$) it corresponds 
to the start of the Wolf-Rayet phase.

Figure~1 also allows us to see the change in relative numbers of BHs
formed through fall back or direct progenitor collapse. This is
important for survival of the BHs in binary systems, since for fall-back 
BHs, which receive natal kicks, the systems hosting BH progenitors may be
disrupted by SN explosions.
For solar metallicity many BHs are formed with kicks through fall back, 
which occurs for single stars with initial masses in the range 
$20-42\,M_\odot$ and $48-70\,M_\odot$. 
For $Z=0.001$, BHs receive a kick in the narrower ranges $18-25\,M_\odot$ and 
$39-54\,M_\odot$.
For an even lower metallicity of $Z=0.0001$, only BHs formed from stars in 
the mass range $18-24\,M_\odot$ receive a kick, while all others form
silently.

\section{RESULTS}

\subsection{Standard Reference Model}

Our reference model starts with $10^6$ 
binaries and $10^6$ single stars (initial binary fraction $f_{\rm bi}=50\%$). 
The initial masses of single stars and binary system primaries (more massive) 
are chosen from a standard initial mass function (IMF) with slope $-2.35$
between $4\,M_\odot$ and the maximum mass $M_{\rm max}$, characteristic of 
young clusters (see Kroupa \& Weidner 2003 for detailed discussion).
We adopt $M_{\rm max} = 150\,M_\odot$ for the reference model calculation. 
The masses of secondary stars in binary systems are sampled from a mass
ratio ($q$, secondary/primary) distribution assumed to be constant between
0 and 1 for our standard model calculation (but values below the
hydrogen burning limit at $0.08\,\msun$ are rejected).
The distribution of initial binary separations is assumed to be constant 
in logarithm between the minimum (such that binary components at ZAMS are 
not in contact) and $10^5\,R_\odot$, while for eccentricities we assume a 
thermal distribution. In our standard model all stars have metallicity 
$Z=0.001$. 

Here we define a BH as a compact object with a mass exceeding the maximum NS mass 
$M_{\rm max,NS}=3\,M_\odot$ (this value will be changed later).
We examine the BHs at five different epochs: 8.7 Myr (corresponding to a turnoff
mass $M_{\rm to} = 25\,\msun$), 11.0 Myr ($M_{\rm to} = 20\,\msun$),  
15.8 Myr ($M_{\rm to} = 15\,\msun$), 41.7 Myr ($M_{\rm to} = 8\,\msun$) and 
103.8 Myr ($M_{\rm to} = 5\,\msun$).
The binaries can produce both BHs in binaries and single BHs.
For the BHs still in binaries we list the numbers formed in various binary
configurations.

Single BHs originating from binaries are formed through several different
channels. 
First, the binary may be disrupted in a SN because of 
mass loss and a possible nascent kick. In this case the evolution of the two 
components is followed as two single stars. Single BHs formed this way will be
listed explicitly under the category "Single: binary disruption".
Binary components may also merge following close interactions 
(e.g., CE evolution, Darwin instability) forming a single
object, which may later evolve and leave behind a single BH 
(denoted by the category "Single: binary merger").
The merger product is assigned a new mass, either the total
component masses for mergers involving compact remnants or MS
stars or their core
masses in cases involving giant-like stars (with the
envelopes assumed lost during the merger process).
If one component is already a BH, we assume that the outcome of the merger is 
still a BH (formed instantaneously, with a new higher mass).
In all other cases, we assume the formation of a new ZAMS star
(i.e., ``full rejuvenation'') and we proceed to evolve the new single star. 
If the merger product is massive enough a (single) BH may form. Since all 
merging binary components are already
evolved, the evolution leading to the formation of a BH in this case 
may in reality be somewhat shorter,
and thus our predicted numbers for this category are only lower limits. 
Most of the merger products massive enough to produce BHs are found to 
have come from pairs of massive HG-MS, He-MS, and MS-MS systems. 

In Table 1 we list the numbers of BHs found both as single objects and as members 
of binary systems. 
Binary BHs are most frequently found either in BH--MS or BH--BH systems.
Shortly after the initial starburst, MS companions naturally dominate 
(they are the most probable BH companions at early times).  
However, as the age increases these massive MS companions themselves 
evolve, collapse to 
BHs and some eventually form BH--BH systems (dominating at later times).
The second longest lived phase in massive star evolution (after the MS) 
corresponds to core helium burning. We note a similar trend for
BH--CHeB as for BH--MS binaries. Although naturally less numerous overall, the 
number of BH--CHeB binaries peaks as soon as these systems appear but then 
tends to decrease with time.  
As the less massive progenitors end their lives and start forming 
remnants, we observe increasing numbers of BH--NS and later BH--WD 
systems. 

There is an overall decrease in the number of binary BHs with time. 
This can be understood as follows. Most massive stars have finished 
their evolution and 
formed BHs at early times ($\la 10$\ Myr). Many of the 
BH binaries with unevolved MS companions either 
merge in RLOF events or are disrupted when the secondary star 
undergoes a SN explosion.
At later times, some of the tightest binaries may also merge due 
to orbital decay driven by GR emission. 
The opposite trend---increase with time---is seen in the total number of 
single BHs. 
After the first $\sim 10$\ Myr most of the single BH progenitors have finished 
their evolution and have formed BHs; therefore the later increase 
is connected to binary evolution. The binary mergers 
and disrupted components continue adding to the single BH population 
at later times.  
For an initial 50\% binary fraction we predict that single BHs 
eventually dominate by about an order of magnitude over BHs found 
in binaries.

There is an interesting direct relationship between the relative numbers of
binary and single BHs for any specific starburst and its initial
(primordial) binary fraction. Etc.
Let us define the ratio of binary to single BHs (as a function of the initial
binary fraction $f_{\rm bi}$) as $R(f_{\rm bi})= (100\%\ N_{\rm bin}) / N_{\rm sin}$, 
where $N_{\rm bin}$  and $N_{\rm sin}$ denote the total number of BHs found in binaries 
and as single objects, respectively. Obviously, this ratio is also a function 
of age, but, for definiteness, let us pick a specific age of $t=11\,$Myr.
For an initial population consisting {\em entirely\/} of binaries ($f_{\rm
bi}=100\%$ from Table~1 we can read that the total number of binary BHs is 27729 
while the number of single BH is 64271 (all from binary disruptions and mergers), 
so the ratio $R(f_{\rm bi}=100\%)=0.37$. This is obviously an upper limit. 
Any contribution from single stars in the initial population will lower this value.

\subsection{Parameter Study and Normalization}

To assess the effects of various model assumptions on BH formation, we
repeat our calculations for a number of different models,
each differing from our standard reference model in
the value of one parameter or one assumption.  
The specific models we chose are described in Table~2.

\emph{Normalization}: In addition to a brief description of each model,
Table~2 also lists for each simulation the total initial  mass of
single and binary stars.
All stars are assumed to form in an instantaneous burst of star formation.  
Stars can form from the hydrogen burning limit up to the 
maximum mass $M_{\rm max}$ characterizing a given system.
We adopt the three-component, power-law IMF of Kroupa, Tout, \&
Gilmore (1993) with slope $\alpha_1=-1.3$ within the initial 
mass range $0.08-0.5\,\msun$, $\alpha_2=-2.2$ for stars within $0.5-1.0\,\msun$, 
and $\alpha_3=-2.35$ within $1.0\,M_\odot - M_{\rm max}$.
This IMF is easily integrated to find the total mass contained 
in single and binary stars for any adopted $\alpha_1, \alpha_2$ values.
The particular choice of low-mass end slope of the IMF ($\alpha_1, \alpha_2$) 
usually (with the exception of Model~B) does not change our results. 
Low-mass stars do not contribute to the BH populations. 
However, as most of the initial stellar mass is contained in low-mass stars, 
a small change in the IMF slope at the low-mass end can significantly change 
the normalization. 
Similarly, our results can easily be generalized to other primordial binary 
fractions ($f_{\rm{bi}}$) by simply weighing differently the results obtained 
for single stars and for binaries. 

\emph{Summary of results}: In Table 3 we summarize the state of each simulation 
at 11 Myr ($M_{\rm to} = 20 \msun$) for the models presented in Table~2. 
In the following, we describe in more detail how the changes made in each model
affect the various BH populations.

In Model~B we create the binary by independently (1) drawing the primary from
$4-150 M_\odot$ according to an IMF exponent -2.35 and (2) drawing the secondary 
from $0.08-150 M_\odot$  via the Kroupa broken power-law IMF. This initialization 
process results in the ratio $q$ of component masses peaked at low $q$ values. 
For this model, almost all BHs in binaries are found with MS companions. In addition, 
relative to the reference model, fewer single BHs are produced from disrupted 
binaries. Since the initial mass ratio is on average smaller in Model~B, 
massive stars, the progenitors of BHs, are formed preferentially with low-mass 
MS companions. If a BH progenitor does not destroy the binary, then the low mass 
companion remains for a long time on MS (large number of BH--MS systems) and it 
does not initiate either RLOF nor explode as SN (small numbers of single BHs).
In contrast to the standard model, there are almost no BH--BH systems, as 
they originate typically from binary progenitors with relatively high 
mass ratios that are almost absent in Model~B.  

For Model~C1 we use small metallicity $Z=0.0001$  (characteristic of 
old Pop II stars metallicity) while  for Model~C2 we choose high (solar) 
metallicity of $Z=0.02$.  We find that the total number of BHs formed can 
vary strongly with the initial chemical abundance. While single BHs are not greatly
affected, the population of binary BHs changes drastically with metallicity. 
With increasing metallicity the evolution of the star is altered; most
notably, the stellar wind mass loss rate increases, causing the
star's mass to drop more rapidly. Eventually, some stars either form lower mass BHs
or instead of forming BHs they end up as NSs. Also higher wind mass loss
rates (Model~C2) and their associated angular momentum losses leave binaries 
wider, and therefore {\em i)} easier to disrupt (see the increase of 
single BHs from binary disruption), {\em ii)} making component interactions 
harder (smaller chance of RLOF; see the smaller merger BH number).
To summarize, then, the BH population formed from binary stars depends
sensitively on the characteristic metallicity of the primordial population.
 
In Model~D we decrease the CE efficiency to 0.1, which means that only 
10\% of binary orbital energy may be used for envelope ejection. 
This leads to either merger of binary components in CE phase, or results in
a much tighter orbit of post-CE binary, if it survives. 
There is no significant change in the number of BHs in Model~D as
compared to the reference model.
Only the close binaries with rather extreme mass ratios undergo a
dynamically unstable RLOF, and thus their evolution may be altered in 
Model~D. This obviously depletes number of BH--MS stars (as they tend to 
have rather small mass ratios, otherwise a MS star would have already evolved 
to a BH), while leaves the number of BH--BH systems unchanged (formed at early
times out of comparable mass binary components). Further orbital shrinkage during 
the dynamically unstable RLOF events obviously increases the number of binary 
mergers (especially in the second RLOF episode), but the merger 
products do not have enough time to form BHs after 11 Myrs. \footnote{Note 
that at later epoch the number of binary mergers has increased by factor of
$\sim 3$ (see Table 4).}  
Finally the binaries which evolved through one or two CE events are tighter 
and are harder to disrupt in SN explosions (explaining the
 slight decrease in the number 
of BHs from binary disruptions). 

In Model~E, we apply for all BHs, independent of their mass and formation
scenario, full nascent kicks following the distribution of Arzoumanian et al.\ (2002). 
By contrast, in our standard model, we either have used smaller kicks or no 
kicks at all for BH formation.  Therefore, we expect (and observe) a drastic 
decrease in number of BHs formed in binary systems, since full kicks disrupt 
all wide and also many close binaries. Further, since systems that would 
otherwise be binary BHs are disrupted, we produce many more single BHs than 
in our reference model. Although the nascent kicks play an important role in
the modeling of BH binary populations, we note that the most
massive BHs almost certainly form silently (Mirabel \& Rodrigues 2003),
without nascent kicks, 
so the results of Model~E should be treated as an extreme case. 

In Model~F, for high-mass stars (potential BH progenitors), we use a steeper IMF
exponent ($\alpha_3=-2.7$), similar to the one observed in field
populations. As expected, all the numbers for BHs found in binaries and as
single objects are decreased. Although we do not expect the initial starburst 
IMF to be as steep as the one observed in the field (Kroupa \& Weidner 2003), 
it is worth noting the factor of 2 decrease in BH numbers for this choice.

Since smaller, less massive clusters form stars only up to a smaller maximum 
mass, in Models~G1 and~G2, we decrease $M_{\rm max}$ to 50 
and 100 $M_\odot$, respectively.
A decrease in the maximum stellar mass in a given simulation shifts more stars to
lower mass, depleting the number of potential BH progenitors. Therefore
there is a general decrease of BH numbers found in Models~G. Since more BH 
progenitors are removed from initial population in Model~G1, it shows the 
largest change as compared to the standard model calculation. 
However, due to a steep IMF for massive BH progenitors, the high mass 
end missing in Models~G1 and~G2 does not contain many stars and the change 
is not large.   

In Model~H we show results for a lowered maximum mass for NS
formation, which may be as low as $1.8 - 2.3 \msun$ (see, e.g., Akmal, 
Pandharipande, \& Ravenhall 1998).
In this model we adopt $M_{\rm max,NS}=2 M_\odot$, and all compact objects 
over that mass are assumed to be BHs. As expected, this model has slightly more single 
and binary BHs. Clearly the maximum NS mass is not significant compared to 
other model uncertainties. 

Finally, under ``Model~I'' (which is not really a different model), we give the 
numbers of BHs with mass greater than $10 M_\odot$ found in our reference model.  
Since most of the single stars and non-interacting binary components form BHs 
with $M \gtrsim 10 \msun$ (see Fig.~1), there is only a slight decrease in
BH numbers. 

In Table 4 we present the subpopulations of single and binary BHs at 103.8
Myr ($M_{\rm to} = 5 \msun$). The trends are very similar to these observed at
the earlier epoch (Table 3), however, with the increased contribution of 
remnant BH binaries (BH--NS and BH--WD).

\subsection{Orbital Periods of Black Hole Binaries}

\emph{Period distribution for the standard model}: 
In Figure~2 we plot the distribution of orbital periods of BH binaries 
for our standard model calculation at $t=11$\ Myrs 
after starburst. This figure also shows the contributions from the
BH--MS and BH--BH subpopulations. 

Orbital periods of binaries hosting BHs are found in a very wide range 
$P_{\rm orb} \sim 0.1 - 10^6$ days. The distribution is characterized  
by two distinctive peaks, smaller at $P_{\rm orb} \sim 10$ days, and 
larger at $P_{\rm orb} \sim 10^5$ days, with the underpopulated region 
($P_{\rm orb} \sim 10^3$ days) in between. 
The clear distinction between short and long period binaries is caused by 
progenitor system RLOF history or lack of thereof. There is a bifurcation
period at $P_{\rm bur} \sim 10^3$ days: below which systems go at least 
through one RLOF interaction leading most frequently to the orbital 
shrinkage, while the wider systems never interact, but only lose material 
in stellar winds, and their orbits expand.  
The specific value of $P_{\rm bur}$ is different for each binary, and depends 
on the maximum radii the components may reach in their evolution (set by
initial mass and metallicity). If at any moment of the binary evolution 
radius of either component increases over its Roche lobe radius (set by the
period and mass ratio) then RLOF starts. Therefore, the $P_{\rm bur}$ is a
function of a number of parameters describing the binary and its components 
and it depends also on the orbit evolution (e.g., tidal interactions), and it is
found within a range: $P_{\rm bur} \sim 100-10^4$\ days for binary BH progenitors.  
Since the tighter binaries evolve through one or more RLOF interactions,
many of them merge, and single objects are formed, causing the depletion of 
the short-period binary BH population (smaller peak).
Although wide binaries avoid mergers, they are much more prone to the disruption 
in SN explosions. However, since most of BHs are formed  silently or with rather 
small kicks, the depletion is not large, and the peak at long periods is 
substantially higher than for tighter binaries.

The BH--BH binaries  evolve only through GR.
Their orbits decay slowly, and for the tightest binaries
the two components can merge and produce a single BH  (within a Hubble
time).   

Systems containing a non-remnant companion (e.g., BH--MS, BH--CHeB), given enough 
time, will populate other BH subpopulations. In particular, if the companion is
massive enough to form a second BH in the system, and the binary survives
potential interactions and SN explosion, a BH--BH binary will be formed. 
It is worth noting that the majority of the binaries are quite ``soft,'' with long
periods $P_{\rm orb} \gtrsim 100$ days, so that, if placed in a dense stellar
environment (e.g., a globular cluster) they will be easily disrupted,
further enhancing the single BH population. On the other hand, a smaller but
significant number of binaries are hard 
and, through dynamical interactions with other stars, may even become harder 
(evolve to shorter periods).
The overall shape of the distribution remains mostly unchanged at different
times (listed in Table~1). However, as discussed in \S\,3.2, 
the contribution of the BH--BH binaries increases, while the BH--MS systems 
are relatively less frequent at the later times. 

\emph{Dependence of period distribution on model assumptions}: 
In Figure~3 we present period distributions of BH binaries for all our 
alternative evolutionary models. 

In most  cases (e.g. Models~B,~C1,~D,~F,~G1,~G2 and~H), the general 
two-peaked shape of our reference model distribution is preserved. 
For some of these models position and/or relative hight of 
the two peaks may change slightly, for reasons that can be easily
explained in terms of the different model assumptions.
For example in Model~D, we find that the number of tight binaries in
comparison with long-period systems is much smaller than for the reference  
model. Moreover, the gap between the two peaks is more pronounced and wider 
than for the reference model. Model~D simulations used  smaller
CE efficiency,  affecting tight interacting binaries by
shifting them to smaller orbital periods (wider gap) or leading to component
merger (smaller peak at low periods).

For a few models, the changes in the distribution are much more pronounced. 
In Models~C2 (high metallicity) and I (high mass BHs only) the distribution
is peaked at large orbital periods, while the short period BH binaries, are 
relatively very infrequent. For example, the Model~I evolution (which favors
only the highest mass progenitors) leads to a selection and survival of only 
the widest binaries. If the high mass stars, progenitors of the most massive 
BHs, are placed on the tight orbits, they merge in the first RLOF
interaction, and the formation of the tight BH system is aborted.  
In contrast with previous two, in Model~E, we find that almost all of the BH
binaries have short orbital periods. With full kicks applied to all BHs,
almost all wide systems are disrupted.

\subsection{Black Hole Masses}

\emph{Mass distribution for the standard model}:
In Figure~4 we plot the mass distribution of the entire BH population formed in 
our standard Model~A. Additionally, we plot separately the single 
and binary BH subpopulations. 
The distribution shows three well defined peaks, first at $M_{\rm BH} \sim
6-8 \msun$, second at $M_{\rm BH} \sim 10-16 \msun$, and third at $M_{\rm BH} 
\sim 22-26 \msun$, and then it steeply falls off with the increasing BH
mass.
The distribution may be divided into two major contributions, one from the 
population of single BHs, and the other from BHs in binaries. 
We see that single BHs dominate the population (see also \S\,3.2) 
and basically set the shape of the overall mass distribution.
A majority of the single BHs originate either from single star progenitors
or from disruptions of wide (non-interacting) binary stars in SN explosions. 
Therefore, the shape of distribution for single BHs is mainly determined by 
a combination of the initial-to-final mass relation for single BHs (Fig.~1) 
with the IMF.
And, in fact, in Figure~1 (middle panel) we can see the pile up of BH remnants 
corresponding to the three peaks in the BH mass distribution of Figure~4. 
Stars over $\sim 50 \msun$ form $10-16 \msun$ BHs corresponding to the
second, the largest, peak in Figure~4. Stars with initial
masses of $25-35 \msun$ form $\sim 25 \msun$ BHs, providing a majority of
the third peak in the BH mass distribution. 
Finally, the stars with initial masses of $40-50 \msun$ tend to form BHs
with masses of $\sim 7 \msun$, the first peak of the Figure~4 distribution. 
The initial-to-final mass relation leading directly to the
shape of the BH mass distribution is described in detail in \S\,3.1.
The binary evolution may increase or decrease the mass reservoir for
BH formation through RLOF interactions between system components. 
However, with one or two exceptions, the overall shape of the mass distribution 
for single and binary BHs is rather similar. 
Specifically, the most apparent change is that for binary BHs the first narrow 
peak for BH masses of $6-8 \msun$ is missing. 
This peak for single BHs comes from the 
turnover in the initial-to-final mass relation, resulting in a pile
up of BHs in this specific mass range. This characteristic feature of
the initial-to-final mass relation corresponds to a very sharp transition in single
star evolution, from H-rich to naked helium stars, which is caused 
by wind mass loss and the effective envelope removal for single stars above
a certain initial mass. 
Since, for binary stars, removal of the envelope is allowed not only through
stellar winds but also through RLOF interactions, it is allowed for the
entire mass range, and therefore there is no sharp transition between
evolution of H-rich and helium reach stars, and the aforementioned peak 
disappears. 
 
One more important effect of binary evolution on BH final mass is illustrated in
Figure~5, where we show single BHs up to the highest formed BH mass. Single
BHs form either from single progenitors, from components of
disrupted binaries, or through binary mergers. Clearly
we can see that single star progenitors form BHs only up to $\sim 30 \msun$ 
(as expected from Fig.~1). Slightly higher BH masses ($\sim
40 \msun$) are obtained through binary disruption, as some of the 
progenitors may have been rejuvenated in RLOF event to a higher mass in the
preceding binary evolution. All of the most massive BHs, up to $\sim 80 
\msun$, are formed through binary mergers. 
Most mergers are  formed out of HG--MS pairs, due to the very
rapid expansion in the Hertzsprung Gap, leading very often to dynamically
unstable RLOF (in which case we always assume a merger; see Ivanova \& Taam
2004). In case of such a merger, the final mass of the newly
formed single star is the sum of the MS star mass and the core mass of the HG star
(the envelope of the HG star, containing $\gtrsim 50\%$ of the entire star 
mass, is assumed to be
lost in the merger process). Since we allow for significant mass
loss in the merger process, the final merger remnant mass may be underestimated. 
Had we assumed that all mass in the system remains after merger, some of BHs 
produced through this channel would have even larger masses, $\gtrsim 100 \msun$. 
These objects have high enough masses to be classified observationally 
as ``intermediate-mass'' BHs. However, it is important to note that they are 
formed through ordinary binary star evolution, without the need for 
additional dynamical interaction processes.

\emph{Effects of model assumptions on mass distribution}:
In Figures~6 and~7 we present the BH mass distributions for all the models 
in our parameter study.
The overall distribution for the entire BH population is not greatly affected
by different choices of parameter values, with the exception of
metallicity (see Fig.~6). This is easily understood, as the highest mass BHs
are formed only at the low metallicity (Models~C1 and~A) and the lightest BHs 
are formed at the high metallicity (Model~C2) as discussed in \S\,3.1.
Single BH masses (Fig.~7) in many models reach very high values around 
$80\,\msun$, and this is a robust result of our calculations for a number of
different evolutionary and initial conditions. The highest maximum BH masses 
are found for lowest metallicity environments, in a larger systems (with high 
$M_{\rm max}$), and for binaries formed with flat mass ratio distributions,
quite independent of other evolutionary parameters.
Only in a few models, with high metallicity (~C2), uncorrelated initial binary 
component masses (B), and low $M_{\rm max}$ (G1, G2), does the maximum BH mass 
stay below $\sim 50 \msun$. 

\emph{GR emission}:
BH--BH binaries (rather than double NSs), are probably
the best candidates for detection by ground-based 
interferometers (Lipunov, Postnov, \& Prokhorov 1997; Bulik \& Belczynski 
2003). BH--BH systems therefore are an important candidate for present  
projects to detect astrophysical GR sources (LIGO, VIRGO). 
The properties of BH--BH binaries at different metallicities formed 
within much larger stellar systems with continuous star formation (e.g., Galactic 
disk) were extensively studied previously (Bulik \& Belczynski 2003; Bulik, Belczynski 
\& Rudak 2004a; Bulik, Gondek-Rosinska \& Belczynski 2004b). We find that the
properties of BH--BH binaries in starbursts are not too different from those
found in the previous studies. Most of BH--BH systems are characterized by
rather equal masses, with a mass ratio distribution peaking at $q \simeq 0.8-1.0$ 
(for comparison see low-metallicity models of Bulik et al.\ 2004b).
For most models only a small fraction ($\sim$ few per cent; e.g., 2\% for Model~A) 
of the BH--BH systems are tight enough to merge within a Hubble time and produce 
observable GR signals. However, for models which tend to produce tighter BH--BH
systems (D, E) the fraction can be significantly higher ($\sim$ 10--20 \%).

\section{SUMMARY}

We have calculated the evolution of the massive stars found in 
young stellar environments, such as starbursts.
The final products of massive star evolution, single and binary BHs, were then
studied. 
All our results were discussed taking into account the many model uncertainties.
A number of alternative calculations with varied initial conditions  
and evolutionary parameters were performed and presented.
We also supplied the necessary data to normalize our 
results to any given total mass of a starburst galaxy or star cluster, 
with arbitrary choice of initial binary fraction. The 
calibrated results may then be used as part of initial conditions for realistic 
$N$-body simulations of dense stellar clusters that include primordial BHs. 

Soon after the initial star formation burst, most BHs are found as single
objects, although a significant fraction of BHs are also found in binaries. 
A number of single BHs are formed as the end product of binary evolution,
either through binary disruption following a SN explosion or through a
merger following a dynamically unstable RLOF episode. The most common 
binary BHs are  BH--MS and BH--BH systems. 
The period distribution of binaries containing BHs is usually bimodal, with
a majority of systems in the long period peak ($P \sim 10^4 - 10^6$\ days). 
These wide binaries would be ``soft'' if placed in a dense stellar environment 
(e.g., a globular cluster) and they would then be disrupted following any strong 
interaction with another passing star or binary, thereby further enhancing the 
population of single BHs. The remaining, short-period BH binaries ($P \sim 
1-100$\ days) would instead undergo hardening and evolve, over many relaxation 
times, to produce a population of very compact binaries that could eventually 
merge through GR emission.

The typical BH masses are found to be within the range $7-25\,\msun$, both for 
single and binary BHs. However, the single BHs formed through binary mergers can
reach masses as high as $\sim 80 \msun$. Since most mergers are assumed in our models
to be accompanied by significant mass loss, BHs formed through binary evolution 
(without any dynamical interactions) could in principle reach even 
higher masses, up to 
$\sim 100 \msun$ (in the absence of significant merger-induced mass loss).
This result has many important implications. 
First, some ultra-luminous X-ray sources might be explained by a $\sim 100 
\msun$ {\em stellar\/} BH accreting from a lower-mass companion. This would 
require the capture of a new companion (most likely through an exchange interaction
with another binary), but no dynamics would be involved in the 
BH formation (cf.\ Kalogera, King, \& Rasio 2004). 
Second, these most massive stellar BHs may act as seeds for the formation of 
true IMBHs (with masses $\gtrsim 1000 \msun$) that could reside at the centers 
of some dense star clusters (Gebhardt, Rich, \& Ho 2002; Gerssen et al.\ 2002;
Miller \& Hamilton 2002). 
Third, a broader mass range for the tightest BH--BH binaries (possibly undergoing 
further hardening through dynamical interactions in a dense star cluster) will 
modify predictions for the gravitational-wave signals detectable by 
laser-interferometer instruments such as LIGO and VIRGO (Flanagan \& Hughes 1998).

\acknowledgements 
We thank A.\ G\"urkan, R.\ O'Shaughnessy and M.\ Tegmark for useful discussions.
FAR acknowledges support from NSF Grants PHY-0133425 and PHY-0245028, 
and from NASA ATP Grant NAG5-12044, and also thanks the Kavli Institute for 
Theoretical Physics for hospitality. KB and AS acknowledge support 
from KBN Grant 5P03D01120.

\pagebreak

\begin{deluxetable}{lccccc}
\tablewidth{480pt}
\tablecaption{Black Hole Populations -- Standard Model}
\tablehead{ Type\tablenotemark{a} & 8.7 Myr & 11.0 Myr & 15.8 Myr & 41.7 Myr & 103.8 Myr \\
 & $M_{\rm to}=25\msun$ & $M_{\rm to}=20\msun$ & $M_{\rm to}=15\msun$ & 
$M_{\rm to}=8\msun$ & $M_{\rm to}=5\msun$ }
\startdata
BH--MS&                            17315 &        16207 & 12215  & 6571   & 4004   \\   
BH--HG&                               22 &           16 &    14  & 16     & 7      \\
BH--RG&                                0 &            0 &    0   & 1      & 0      \\
BH--CHeB&                           1254 &         1029 &    675 & 262    & 155    \\ 
BH--AGB&                              16 &           13 &   27   & 9      & 16     \\
BH--He&                              167 &          102 &     60 & 0      & 0      \\
BH--WD&                                0 &            0 &    0   & 1      & 1075   \\
BH--NS&                               69 &          364 &    760 &  913   &  880   \\
BH--BH&                             9261 &         9998 &  10022 & 10010  & 9996   \\
Total in binaries:&               28104 &        27729 &  23773 & 17783  & 16133  \\ 

&&&&&\\

Single: binary disruption&        24093 &        43909 &  55649 & 60262  & 61021  \\
Single: binary merger&            10128 &        20362 &  33825 & 65236  & 66148  \\  
Single progenitor&                77580 &       108080 & 120100 & 120100 & 120100 \\  
Total single:&                    111801&       172351 & 209574 & 245598 & 247269 \\ 

\enddata
\label{std}
\tablenotetext{a}{Black holes in binary systems are listed according to their
companion types: MS---main sequence, HG---Hertzsprung Gap, RG---reg giant, CHeB---core 
He burning, AGB---asymptotic giant branch, He---helium star, WD---white dwarf,
NS---neutron star, BH---black hole.
Single black holes formed from components of disrupted binaries are listed
under ``Single: binary disruption.''  
Single black holes formed from binary merger products are under 
``Single: binary merger.'' 
Single black holes that are remnants of single stars are listed under ``Single
progenitor.''
}
\end{deluxetable}

\begin{deluxetable}{llcc}
\tablewidth{390pt}
\tablecaption{Population Synthesis Model Assumptions}

\tablehead{ 
      &                              & Mass [$M_\odot$] in  & Mass [$M_\odot$] in \\
Model & Description\tablenotemark{a} & Single Stars & Binaries }
\startdata
A      & standard model described in \S\,3.1    & $3.8 \times 10^{7}$ & $5.9 \times 10^{7}$ \\
B      & uncorrelated binary component masses   & $3.8 \times 10^{7}$ & $7.6 \times 10^{7}$ \\
C1-2   & metallicity $Z = 0.0001, 0.02$         & $3.8 \times 10^{7}$ & $5.9 \times 10^{7}$ \\
D      & $\alpha_{\rm CE} \times \lambda = 0.1$ & $3.8 \times 10^{7}$ & $5.9 \times 10^{7}$ \\
E      & full kicks for BHs                     & $3.8 \times 10^{7}$ & $5.9 \times 10^{7}$ \\
F      & steeper IMF: $\alpha_3=-2.7$         & $6.0 \times 10^{7}$ & $9.5 \times 10^{7}$ \\ 
G1     & lower maximum mass: $M_{\rm max} = 50 M_\odot$  & $3.7 \times 10^{7}$ & $5.8 \times 10^{7}$ \\
G2     & lower maximum mass: $M_{\rm max} = 100 M_\odot$ & $3.8 \times 10^{7}$ & $5.9 \times 10^{7}$ \\
H      & $M_{\rm max,NS}=2 $\,M$_\odot$         & $3.8 \times 10^{7}$ & $5.9 \times 10^{7}$ \\ 
I\tablenotemark{b} & BHs more massive than 10 M$_\odot$ & $3.8 \times 10^{7}$ & $5.9 \times 10^{7}$ \\ \\
\enddata
\label{models}   
\tablenotetext{a}{Details of model assumptions are given in 
\S\,3.2 and \S\,3.3.}
\tablenotetext{b}{Model~I is shown only to give the numbers of 
 BHs (formed in the standard Model~A) with mass greater than 
 $10 M_\odot$.}
\end{deluxetable}

\vspace*{10cm}
\begin{deluxetable}{lccccccccccc}
\tablewidth{540pt}
\tablecaption{Very Young Black Hole Populations -- Parameter Study\tablenotemark{a}}
\tablehead{ Type&         A & B & C1 & C2 & D & E & F & G1 & G2 & H & I }
\startdata
Binaries:&&&&&&&&&&&\\
BH--MS&                   16207 & 32464 & 25527  &  2831  &  12683 &  3176  & 9656   & 13434 & 15798  & 17566  & 11024  \\
BH--HG&                      16 &     1 &    26  &     7  &    25  &     6  &   10   & 17    &    28  &    16  &    10  \\
BH--RG&                       0 &     0 &     0  &     0  &     0  &     0  &    0   & 0     &     0  &     0  &     0  \\
BH--CHeB&                  1029 &    16 &  1571  &   200  &    967 &    58  &  616   & 776   &  1032  &  1047  &   894  \\
BH--AGB&                     13 &     0 &    20  &     7  &   25   &     0  &   10   & 13    &    20  &    13  &    12  \\
BH--He&                     102 &     3 &   204  &    71  &    26  &    76  &   64   & 102   &   116  &   106  &    22  \\
BH--WD&                       0 &     0 &     0  &     0  &    0   &     0  &    0   & 0     &     0  &     0  &     0  \\
BH--NS&                     364 &     2 &   545  &    94  &    99  &   106  &  180   &  305  &   364  &    55  &     0  \\
BH--BH&                    9998 &    26 & 20914  &   2743 &  10129 &   196  & 4784   & 3127  &  7496  &  10326 &  7674  \\
Total:&                  27729 & 32512 & 48807  &   5953 &  23954 &   3618 & 15320  & 17774 & 24854  & 29129  & 19636  \\        

&&&&&&&&&&&\\

Single:&&&&&&&&&&&\\
binary disruption&       43909 &   18929 &  21037 &  38110 &  34806 &  79260 & 25696 &  35858 &  42022 &  47714 &  10863 \\
binary merger&           20362 &    7402 &  16790 &  33809 &  20084 &  19830 & 10552 &  10837 &  18692 &  20793 &  18030 \\
Single progenitor&      108080 &  108080 & 109755 & 122470 & 108080 & 108080 & 63090 &  83695 & 102255 & 108080 &  87630 \\
Total:&                 172351 &  134411 & 147582 & 194389 & 162970 & 207170 & 99338 & 130390 & 162969 & 176587 & 116523 \\

\enddata
\label{parm}
\tablenotetext{a}{All numbers correspond to an age of 11 Myrs ($M_{\rm to}=20 M_\odot$)}
\end{deluxetable}

\vspace*{10cm}
\begin{deluxetable}{lccccccccccc}
\tablewidth{540pt}
\tablecaption{Young Black Hole Populations -- Parameter Study\tablenotemark{a}}
\tablehead{ Type&         A & B & C1 & C2 & D & E & F & G1 & G2 & H & I }
\startdata
Binaries:&&&&&&&&&&&\\
BH--MS&                        4004   & 31762 & 5952   & 888   & 3052  & 766   & 2520  & 3394  & 3998       & 4506  & 2836 \\ 
BH--HG&                        7      & 2     & 14     & 0     & 11    & 0     & 6     & 5     & 14         & 7     & 5 \\
BH--RG&                        0      & 0     & 0      & 1     & 0     & 0     & 2     & 0     & 0          & 0     & 0 \\
BH--CHeB&                      155    & 50    & 238    & 55    & 153   & 2     & 96    & 104   & 164        & 155   & 151 \\
BH--AGB&                       16     & 3     & 22     & 6     & 18    & 0     & 8     & 16    & 14         & 16    & 15 \\
BH--He&                        0      & 0     & 0      & 0     & 0     & 0     & 0     & 0     & 0          & 1     & 0 \\
BH--WD&                        1075   & 205   & 1470   & 459   & 1064  & 10    & 612   & 779   & 966        & 1087  & 1033 \\
BH--NS&                        880    & 13    & 1705   & 160   & 292   & 356   & 512   & 763   & 980        & 610   & 56 \\
BH--BH&                        9996   & 26    & 20915  & 2743  & 9745  & 202   & 4790  & 3130  & 7494       & 10434 & 7674 \\
Total:&                       16133  & 32061 & 30316  & 4312  &14335  & 1336  & 8546  & 8191  & 1630       & 16816 & 11770 \\

&&&&&&&&&&\\

Single:\\
binary disruption&            61021  & 23660 & 44889  & 40146 & 51089 & 88394 & 36806 & 51708 & 58766      & 88195 & 15167 \\
binary merger&                66148  & 19012 & 54904  & 88961 & 74952 & 64172 & 43794 & 57573 & 65004      & 68834 & 49515 \\
Single progenitor&            120100 &120100 & 147065 &127835 &120100 &120100 & 93080 &124755 & 142765     & 148465& 87630\\
Total:&                       247269 &162772 & 246858 &256942 &246141 &272666 &173680 &234036 & 266535     & 305494& 152312 \\

\enddata
\label{parm1}
\tablenotetext{a}{All numbers correspond to an age of 103.8 Myrs ($M_{\rm to}=5 M_\odot$)}
\end{deluxetable}

\psfig{file=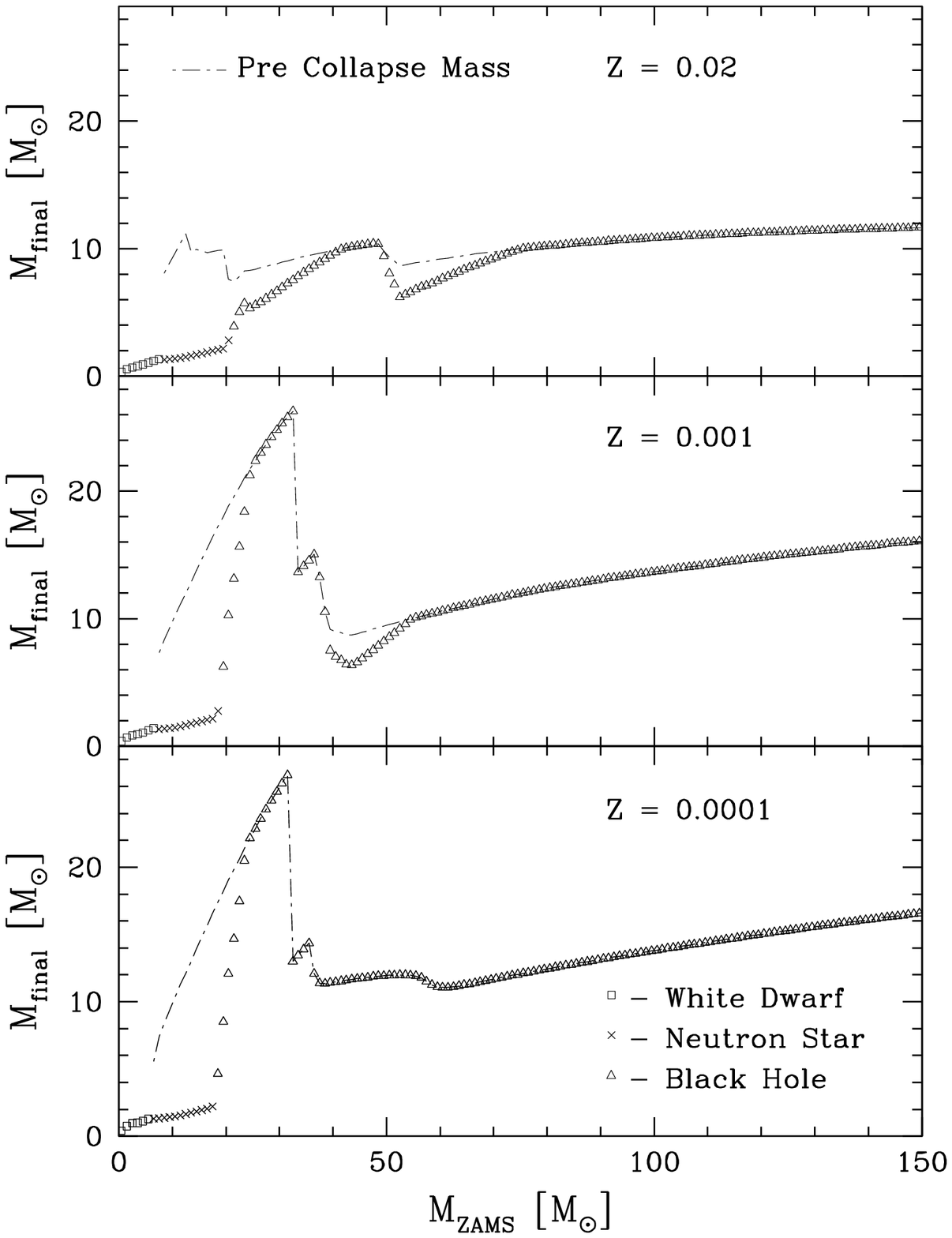,width=1.0\textwidth}
\figcaption[]{
\footnotesize
Initial-to-final mass relation for different metallicities.
Remnants of different type: white dwarfs (WD), neutron stars (NS) and black
holes (BH) are marked.
}

\psfig{file=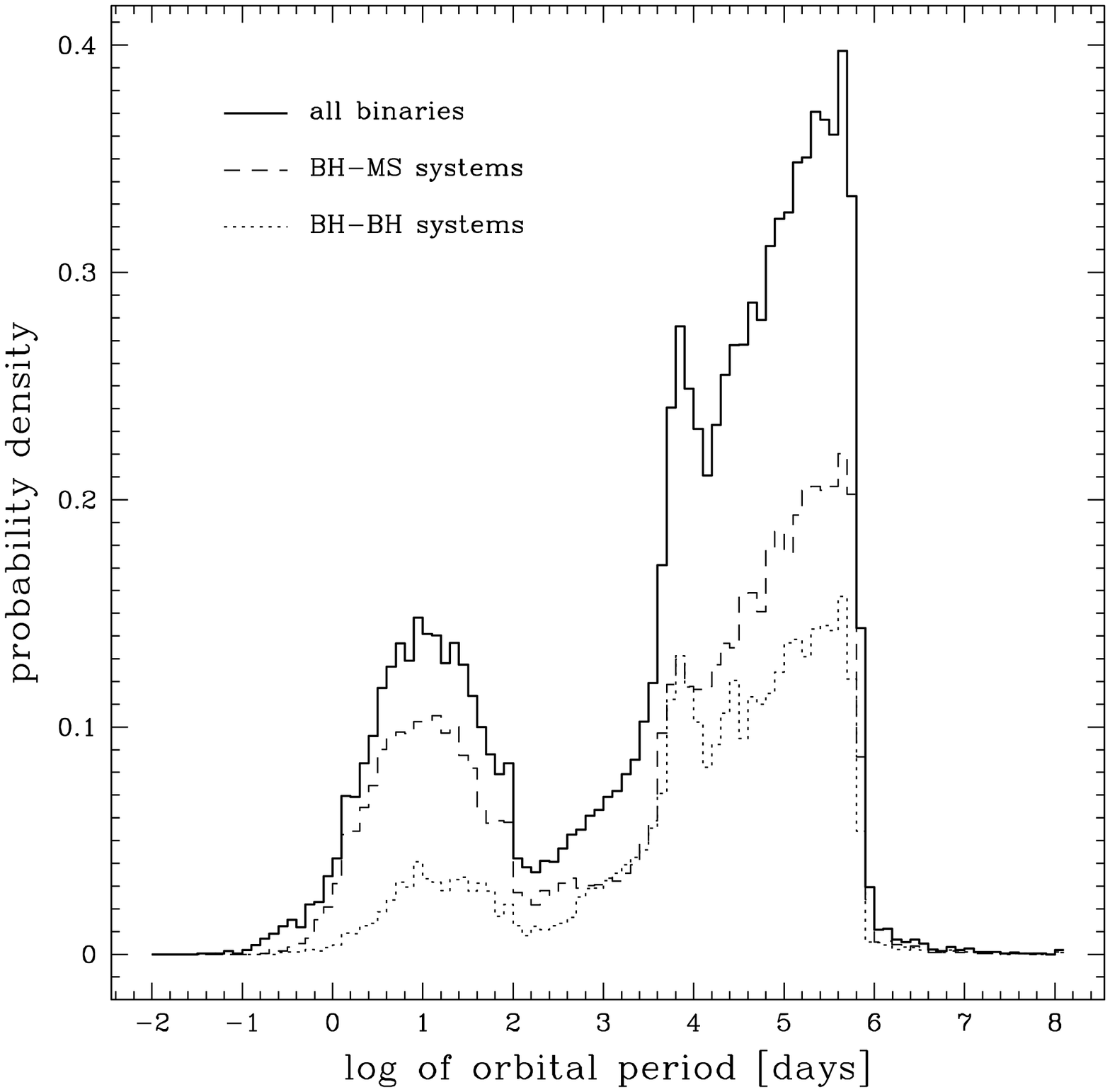,width=0.95\textwidth}
\figcaption[]{
\footnotesize
Period distribution of BH binaries (solid line) for our standard model
after 11 Myr.
Two major contributing system types are shown separately: BH--MS binaries 
(dashed line) and  BH--BH binaries (dotted line).
Normalized to total number of BHs (single and binaries), binwidth: tenth of
decade.
}

\psfig{file=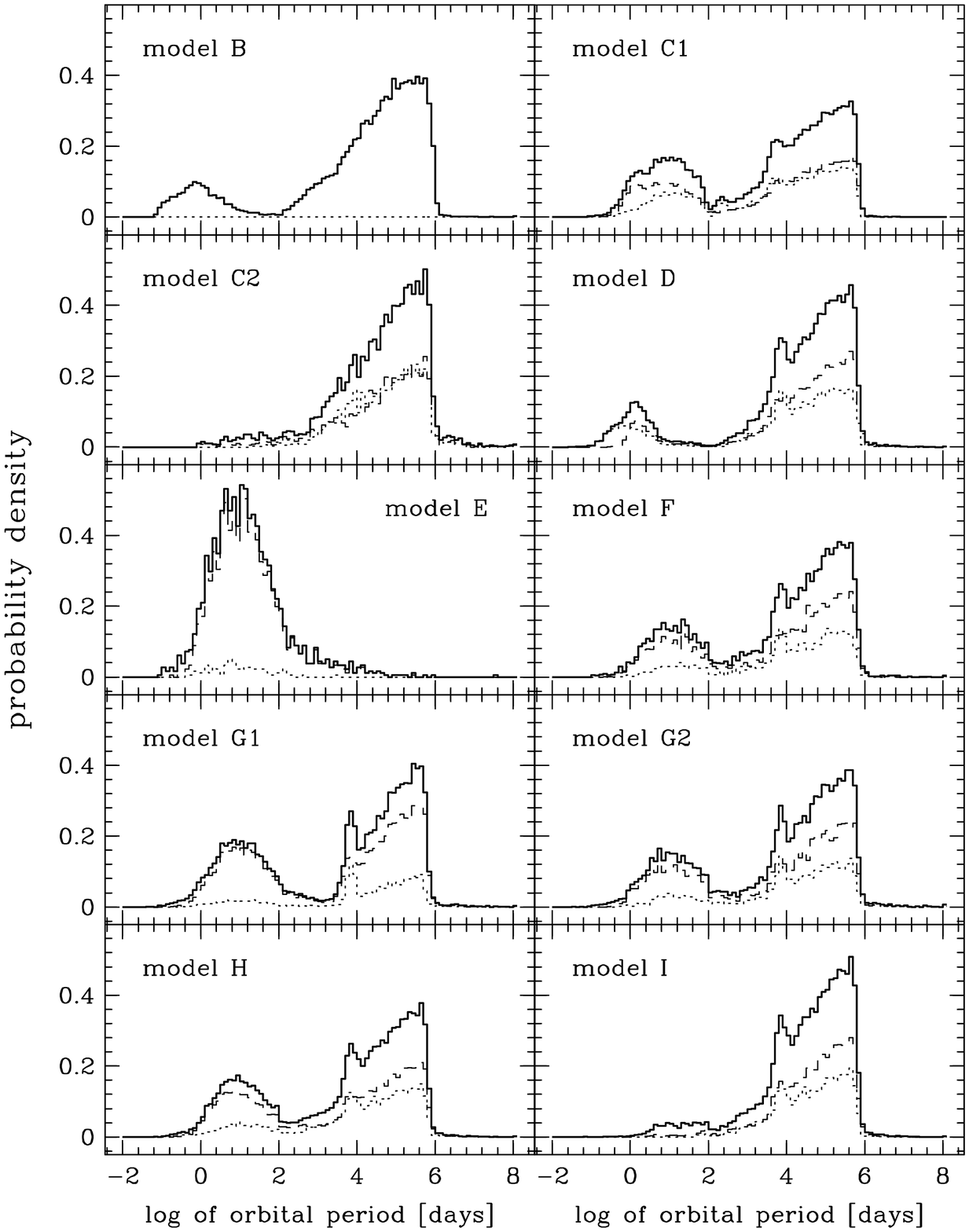,width=0.95\textwidth}
\figcaption[]{
\footnotesize
Period distribution of BH binaries (solid line) for different models,
all at 11 Myr (dashed line: 
BH--MS binaries; dotted line: BH--BH binaries).
}

\psfig{file=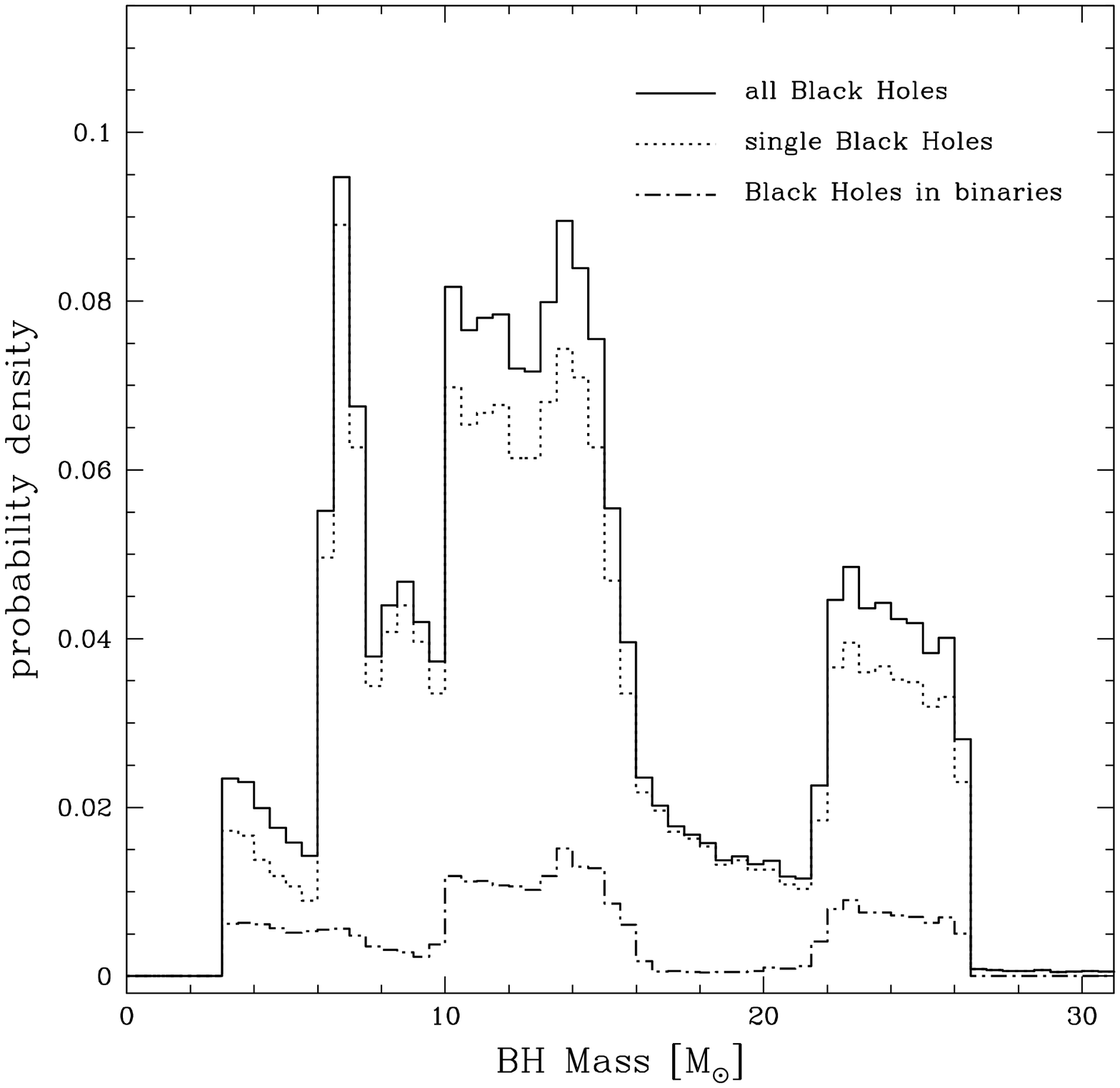,width=0.95\textwidth}
\figcaption[]{
\footnotesize
Mass distribution of BHs at 11 Myr (solid line) for standard model (~A). 
Single BHs are shown with dotted line and BHs in binaries with 
dashed line.
Normalized to total number of BHs (single and binaries); binwidth: 
$0.5 M_\odot$.
}

\psfig{file=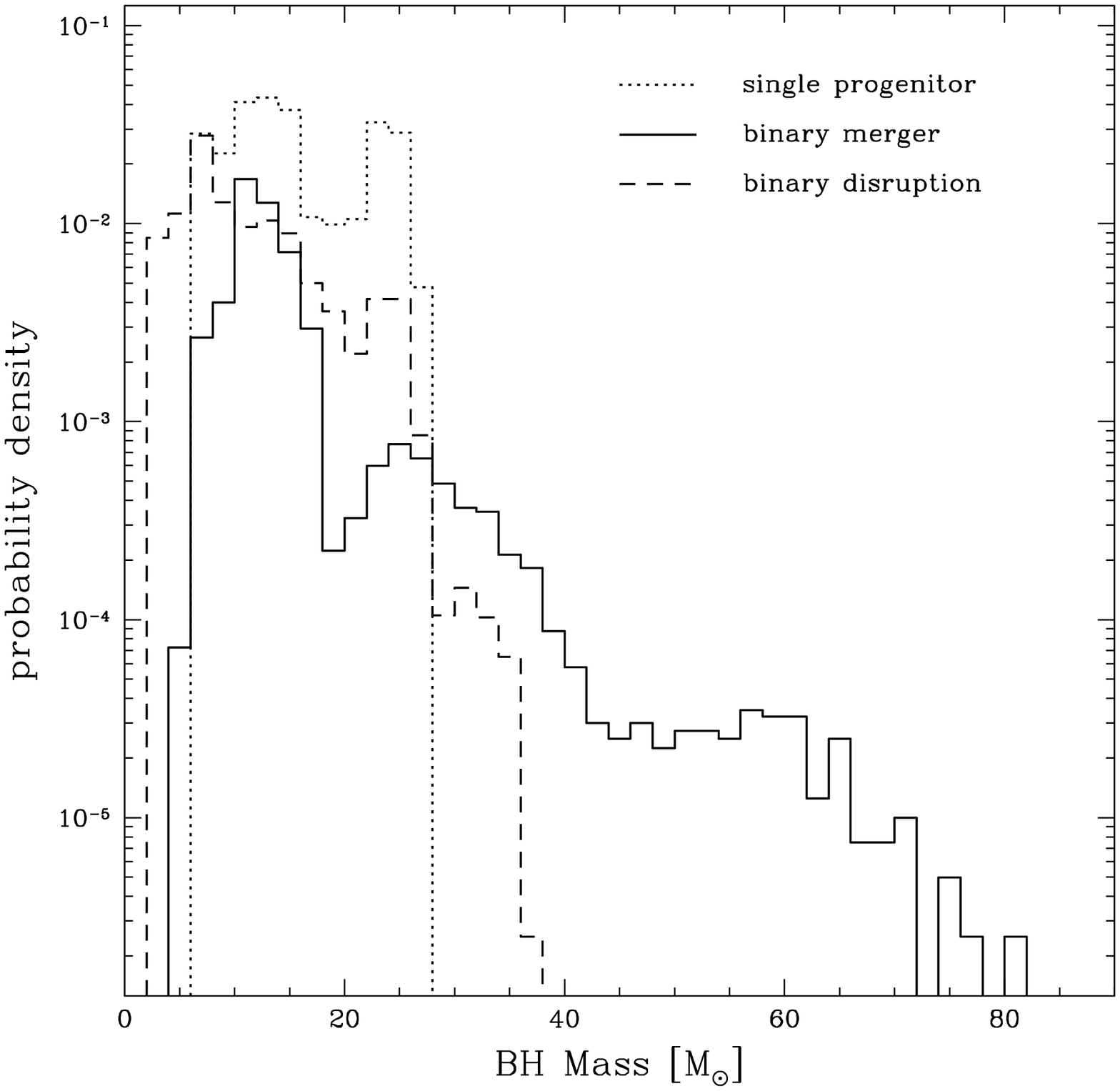,width=0.95\textwidth}
\figcaption[]{
\footnotesize
Mass distribution of various kinds of single BHs at 11 Myr in standard model (~A).
The dotted line shows BHs originating from primordial single stars; the dashed line 
represents single BHs from disrupted binaries; the
solid line is for single BHs that are the remnants of merged binaries. 
Normalized to total number of BHs (single and binaries); binwidth:
$2.0 M_\odot$.
}

\psfig{file=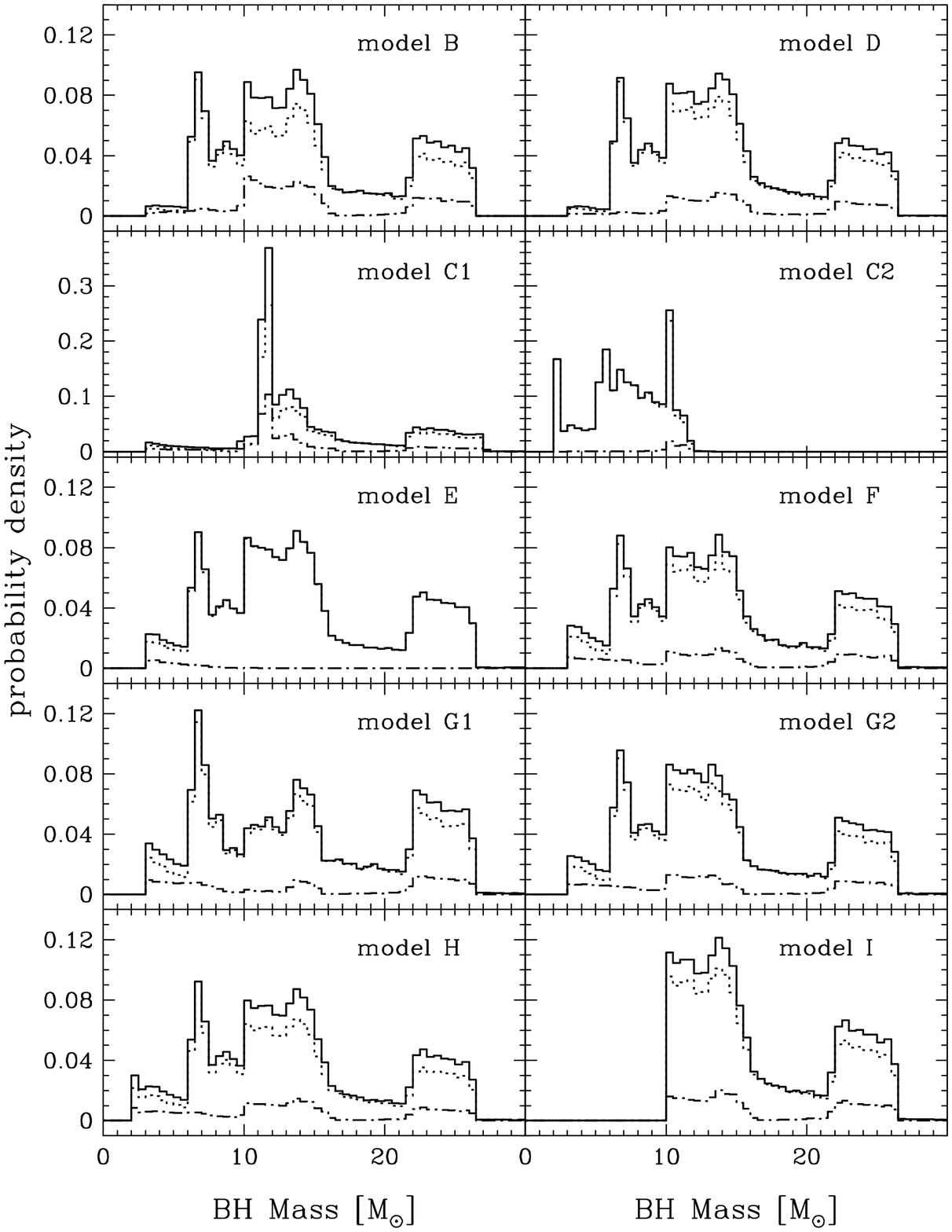,width=0.95\textwidth}
\figcaption[]{  
\footnotesize
Mass distribution of BHs at 11 Myr for different models. 
Conventions are as in Fig.~4.
Note that for Models~C1 and~C2 the vertical scale differs from the
rest of the panels.
}

\psfig{file=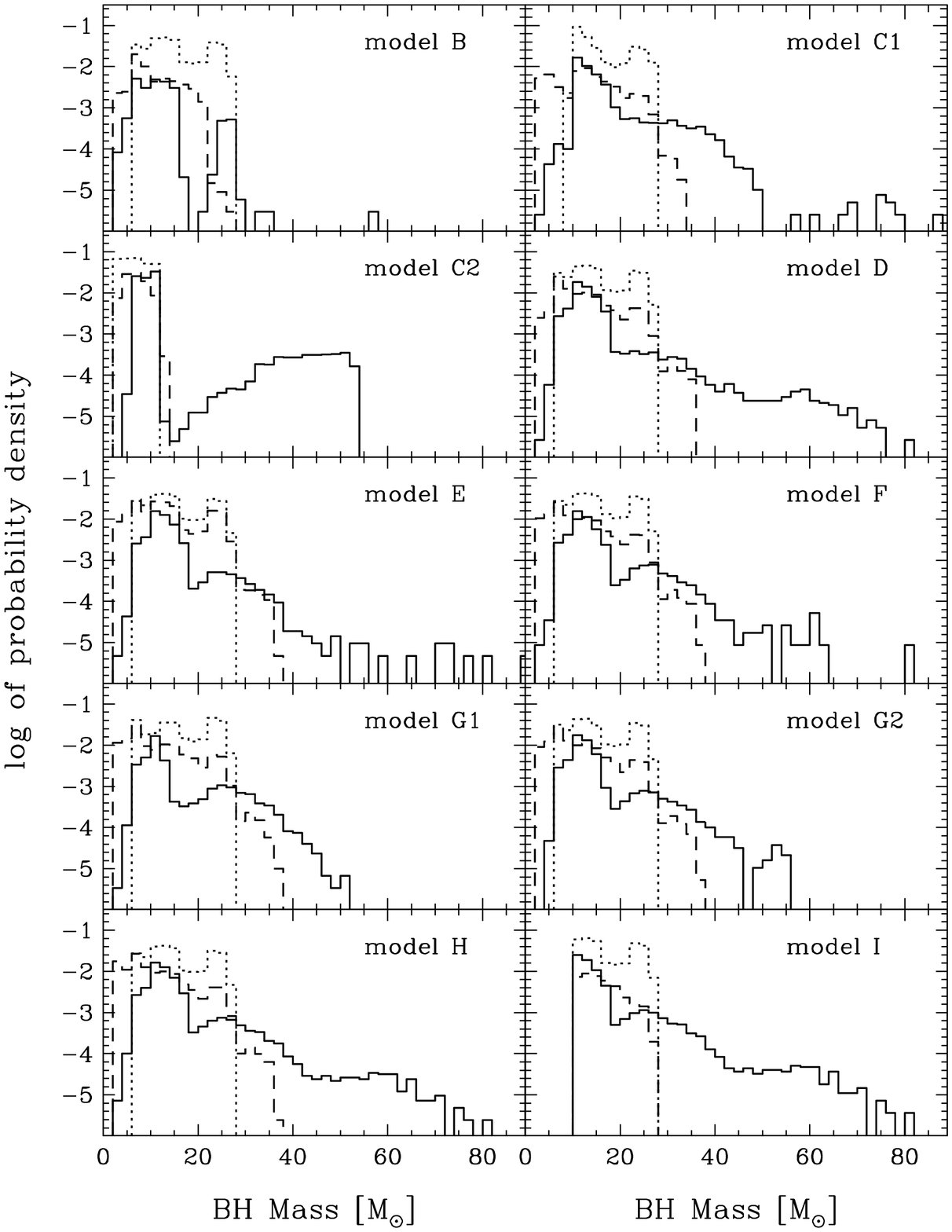,width=0.95\textwidth}
\figcaption[]{
\footnotesize
Mass distribution of various kinds of single BHs at 11 Myr for different models.
Conventions are as in Fig.~5.
}

\end{document}